\documentclass{article}
\usepackage{spconf,amsmath,graphicx}
\usepackage{multirow}
\usepackage{caption}
\usepackage{subcaption}

\title{A Pre-trained Audio-visual Transformer for Emotion Recognition}
\name{Minh Tran \qquad Mohammad Soleymani}
\address{University of Southern California\\
	Department of Computer Science\\
	Los Angeles, USA}

\begin{document}
%
\maketitle
\begin{abstract}
In this paper, we introduce a pretrained audio-visual Transformer trained on more than 500k utterances from nearly 4000 celebrities from the VoxCeleb2 dataset for human behavior understanding. The model aims to capture and extract useful information from the interactions between human facial and auditory behaviors, with application in emotion recognition. We evaluate the model performance on two datasets, namely CREMAD-D (emotion classification) and MSP-IMPROV (continuous emotion regression). Experimental results show that fine-tuning the pre-trained model helps improving emotion classification accuracy by 5-7\% and  Concordance Correlation Coefficients (CCC) in continuous emotion recognition by 0.03-0.09 compared to the same model trained from scratch. We also demonstrate the robustness of finetuning the pre-trained model in a low-resource setting. With only 10\% of the original training set provided, finetuning the pre-trained model can lead to at least 10\% better emotion recognition accuracy and a CCC score improvement by at least 0.1 for continuous emotion recognition. 


\end{abstract}
\section{Introduction}
\label{sec:intro}
Recent advances in machine learning and signal processing enable an unprecedented opportunity to computationally analyze and predict social behaviors. A better understanding of how people behave and express themselves could have wide applicability. Much human interaction research (e.g. emotion recognition) is task-oriented, which often requires time-consuming and expensive data collection processes; and hence, suffers from small population that prevents ML models to generalize well. Despite the scarcity of labeled data, there is an abundance of data on human communication that is unlabeled, multi-modal, and easily accessible \cite{chung2018voxceleb2}. This opens an opportunity to address the challenge, by creating self-supervised pretrained models that are trained on unlabeled data and can be finetuned for downstream tasks. Similar approaches have been very successful in NLP \cite{devlin2019bert} and speech processing \cite{liu2020mockingjay} tasks.

Most research extending the standard Transformer \cite{vaswani2017attention} in a multimodal context focuses on the visual-and-language domain. Existing work generally utilize the language-pretrained BERT \cite{devlin2019bert} and train only the visual components through either the single-stream framework (image and text are jointly processed by a single encoder) \cite{zhou2020unified, zhu2020actbert} or dual-stream framework (with separate visual and text encoders) \cite{tan2019lxmert, lu2019vilbert}. To the best of our knowledge, Lee \textit{et al.} present the only pretrained Transformer-based model for the audio-and-visual domain \cite{lee2020parameter}. Their end-to-end model contains two Transformers to encode audio and visual inputs independently, followed by another Transformer that processes the encoded audio and video signals sequentially. Lee \textit{et al.} pretrain their model on Kinetics-700 (containing 700 human action classes) \cite{carreira2019short} and AudioSet (containing 632 audio event classes) \cite{gemmeke2017audio}. Since both datasets contain little information on human interactions, the pretrained models would not be appropriate for downstream tasks such as emotion recognition.

In this study, we present the first pretrained audio-visual Transformer-based model that learns from human communicative behaviors. We then validate the the pretrained model for the downstream task of emotion recognition on the CREMA-D dataset \cite{cao2014crema} and the MSP-IMPROV dataset \cite{busso2016msp}.

\begin{figure*}[htb]
\centering
\includegraphics[width=0.7\linewidth]{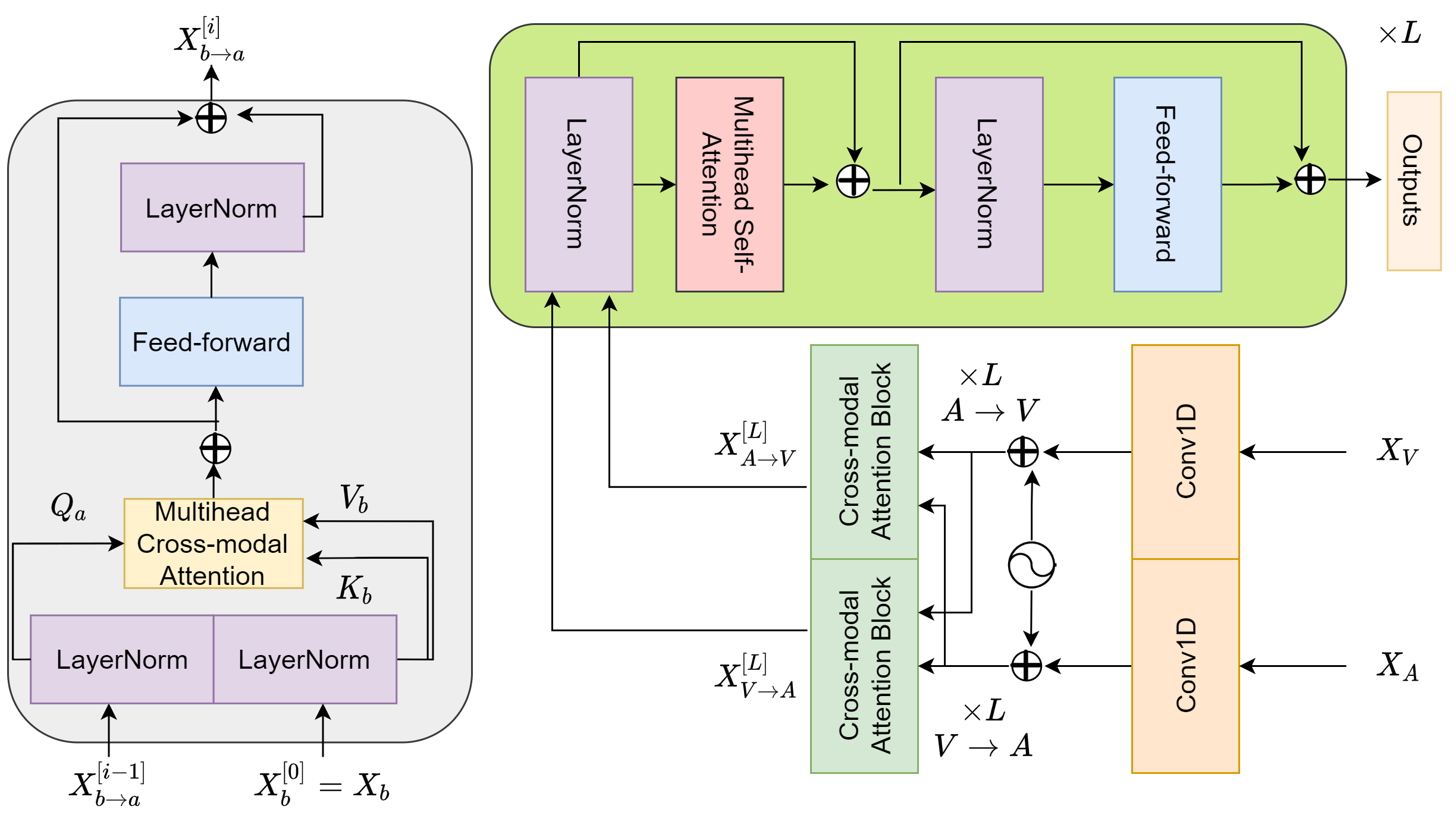}
\vspace{-1em}
\caption{The architecture of the $i^{th}$ layer in a $b\rightarrow a$ Cross-modal Transformer  (Cross-modal Attention Block) is shown on the left. The architecture of the Multimodal Transformer is shown on the right. Re-illustration based on \cite{tsai2019multimodal}}
\label{fig:mult}
\vspace{-10pt}
\end{figure*}

\section{Method}
\label{sec:method}
\subsection{Multimodal Transformer architecture}
We adapt the Multimodal Transformer (MulT) architecture \cite{tsai2019multimodal} for the pretraining task. At the high level, the architecture consists of 4 main components: the temporal convolutions that projects features of different modalities to the same dimension, a sinusoidal positional encoding to capture temporal information, the Cross-modal Transformers that allow one modality to pass information to another, and the standard Self-Attention Transformers that process the fused information produced by the Cross-modal Transformers. An overview of the MulT architecture is available in Figure \ref{fig:mult} (right). 

At the core of MulT is the Cross-modal Attention Block (Fig \ref{fig:mult} left), which differs from the standard Transformer encoder layer in two ways. First, the module merges audio-visual temporal information though the Multihead Cross-modal Attention layer, with the Queries being in one modality while the Keys and Values being in another. Second, each Cross-modal Attention Block learns directly from the low-level feature sequences (\textit{i.e.} $X^{[0]}_b$ is passed to the $b\rightarrow a$ Cross-modal Attention Block regardless of the layer position) while the standard Transformer takes intermediate-level features as input. Tsai \textit{et al.} \cite{tsai2019multimodal} empirically show that adapting low-level features for the Cross-modal Attention Block is beneficial for MulT. 
With only 2 modalities in consideration, we have 2 types of Cross-modal Attention Blocks ($V\rightarrow A$ and $A\rightarrow V$). Eventually, we concatenate the outputs of the Cross-modal Transformers and pass them through a standard Self-attention Transformer \cite{vaswani2017attention}. The outputs of the Self-attention Transformer are finally converted to the original audio and visual feature dimensions for predictions using two independent fully-connected layers.

\subsection{Pretraining procedure}
Following prior work on pretrained Transformers \cite{devlin2019bert, liu2020mockingjay, lee2020parameter}, we use the masked frame prediction task to train our model. Specifically, we randomly select 15\% of the frames, mask them for both the audio and visual inputs, and train the model to reconstruct the masked frames. Following \cite{devlin2019bert, liu2020mockingjay} on the selected frames for masking, we mask the frames all to zero with a probability of $0.8$, replace them with randomly selected frames with a probability of $0.1$ and keep them untouched with a probability of $0.1$. Similar to \cite{liu2020mockingjay}, we use the $L1$-Loss to measure reconstruction error. The loss for each prediction is the sum of the $L1$-loss for the audio modality and the $L1$-loss for the visual modality. We adapt the strategy of masking consecutive frames from \cite{liu2020mockingjay} to prevent the model from exploiting local smoothness. We use dynamic masking for the training set (the masked frames of each input sequence are selected independently every time the sequence is called) and static masking for the validation set (the masked frames for each input sequence are pre-computed) to make the comparison between models' performances fair. 

\subsection{Feature selection}
The Multimodal Transformer is not end-to-end, so we need to extract acoustic and visual features as inputs to the model. Since features can be more or less powerful depending on the context of their usage and the target of our study is emotion recognition, we compare different baseline features on the CREMA-D and MSP-IMPROV datasets \cite{cao2014crema, busso2016msp}. The motivation for pretraining data with MulT is to capture and model temporal dependencies so we also want the base features to be temporally independent. Thus, even though features extracted from pretrained Speech Transformers such as \cite{liu2020mockingjay, liu2021tera, chi2021audio} are powerful, they are not suitable to be base features for MulT. 

With these considerations, we select and compare the features extracted from pretrained Facenet \cite{schroff2015facenet}, pretrained ResNet \cite{he2016deep} and OpenFace Action Units' intensities \cite{baltrusaitis2018openface} for the visual modality. For the acoustic modality, we compare Mel-scale spectrogram, Linear-scale spectrogram and features extracted by TRILL \cite{shor2020towards}. To extract features from FaceNet and ResNet for a given video, we extract frames from the video at a constant rate and crop the face regions before feeding them into the pretrained models. Because TRILL only provides one vector representation for each input acoustic sequence, we split the input sequence into segments that matches a specified frame rate and extract the representations of the segments. To make the comparison fair, we apply all extracted features to the same model (a single-layer Gated Recurrent Unit with a hidden size of $512$ and dropout ratio of $0.2$ with fixed initialization) for emotion classification (CREMA-D dataset) and continuous emotion estimations (MSP-IMRPOV dataset). We find that extracted OpenFace and TRILL features outperform other baseline features by a considerable margin on both datasets. On the CREMA-D dataset, OpenFace shows a gain of $17\%$ in Accuracy in comparison with the ResNet representations, and TRILL shows a gain of more than $11\%$ in Accuracy from the Linear-scale spectrogram. On the MSP-IMPROV dataset, TRILL outperforms the Linear-scale spectrogram with a CCC margin of at least $0.03$ while OpenFace outperforms the second-best baseline with a CCC margin of at least $0.14$. 


\section{Experiments}
\subsection{Data}
\noindent\textbf{Voxceleb2} 
We use the Voxceleb2 dataset for pretraining \cite{chung2018voxceleb2}. It contains more than 1M utterances from more than 6,000 celebrities collected from around 150K videos on Youtube. The dataset is fairly gender balanced (61\% are men). For the acoustic modality, we first segment the audio, into 200ms segments, before feeding them into TRILL \cite{shor2020towards} for feature extraction. Because TRILL originally provides a single embedding for an audio input as a whole, we do not want to extract the features with smaller segment duration. For the visual modality, we use OpenFace2.0 \cite{baltrusaitis2018openface} to track 17 Facial Action Unit (AU) intensities from the videos at 30 FPS. Since there is a high variation in the video quality of the Voxceleb2 dataset, we remove frames with detection confidence below 80\%. We then downsample OpenFace outputs to 5 FPS to match  the frame rate of the acoustic modality. We remove utterance samples with the audio and video features misaligned for more than $1$ second (more than $5$ frames difference). Although MulT can handle unaligned multimodal sequences, the model achieves better performance with aligned sequences. In the end, we end up with a training dataset of 524K utterances from about 4K speakers (the average duration of each utterance is $5$s with a standard deviation of $0.7$s). 

\noindent\textbf{CREMA-D} 
The CREMA-D dataset is an acted audiovisual database consisting of 6 basic emotional states (happy, sad, anger, fear, disgust and neutral). It includes 7,442 video clips from 91 actors speaking 12 sentences with different emotions. The emotion labels are collected through crowd-sourcing from 2,443 raters, and the human recognition accuracy of intended emotion is $63.6\%$. The emotion classes in the dataset are balanced. In this study, we perform speaker-independent split of the CREMA-D dataset into the train-validation-test set with a ratio of 60\%-20\%-20\% respectively. 

\noindent\textbf{MSP-IMRPOV} 
The MSP-IMPROV dataset is an acted audiovisual database that includes emotional interactions between people in a dyadic conversational setting. The conversation scenarios are designed to invoke realistic emotions. The dataset consists of 8,450 video recordings that are recorded during 6 dyad sessions from 12 actors. The annotations for the dataset are collected via crowd-sourcing, and each video is annotated with at least 5 evaluators. The annotation includes emotional content and a five-point Likert-like scale on valence (1-negative and 5-positive), arousal (1-excited and 5-calm) and dominance (1-weak and 5-strong). In this study, we focus on the continuous emotion regression task that estimates the values of Valence and Arousal for a given video. We use Session 1-4 as the training set, Session 5 as the validation set and Session 6 as the test set.  

\begin{table}[]
\centering
\resizebox{\columnwidth}{!}{%
\begin{tabular}{|p{1.8cm}|p{1.5cm}|p{1.1cm}|p{1cm}|p{1.1cm}|p{1cm}|}
\hline
\multirow{3}{*}{} & \multirow{2}{*}{CREMA-D} & \multicolumn{4}{c|}{MSP-IMPROV}                             \\ \cline{3-6} 
                  &                          & \multicolumn{2}{c|}{Arousal} & \multicolumn{2}{c|}{Valence} \\ \cline{2-6} 
                  & Accu. $\uparrow$                & MAE $\downarrow$           & CCC $\uparrow$          & MAE $\downarrow$          & CCC $\uparrow$          \\ \hline
TFN \cite{zadeh2017tensor}              & 63.09                    & 0.466         & 0.581        & 0.596         & 0.592        \\ \hline
EF-GRU            & 57.06                        & 0.676             &  0.399           & 0.774             & 0.478            \\ \hline
LF-GRU            & 58.53                        & 0.496             & 0.546            & 0.619             & 0.579            \\ \hline
MulT WOP & 63.93                    & 0.466         & 0.665        & 0.580         & 0.607        \\ \hline
MulT BASE         & 68.87                    & 0.456         & \textbf{0.697}        & 0.576         & 0.658        \\ \hline
MulT Large        & \textbf{70.22}                    & \textbf{0.431}         & 0.693        & \textbf{0.563}         & \textbf{0.692}        \\ \hline
\end{tabular}}
\caption{Comparison between the performances of different models.  WOP stands for w/o Pretraining. }
\label{tab:comparisons}
\vspace{-1em}
\end{table}

\begin{figure}[h!t]
\centering
\begin{subfigure}[b]{\columnwidth}
\centering
  \includegraphics[width=.8\linewidth]{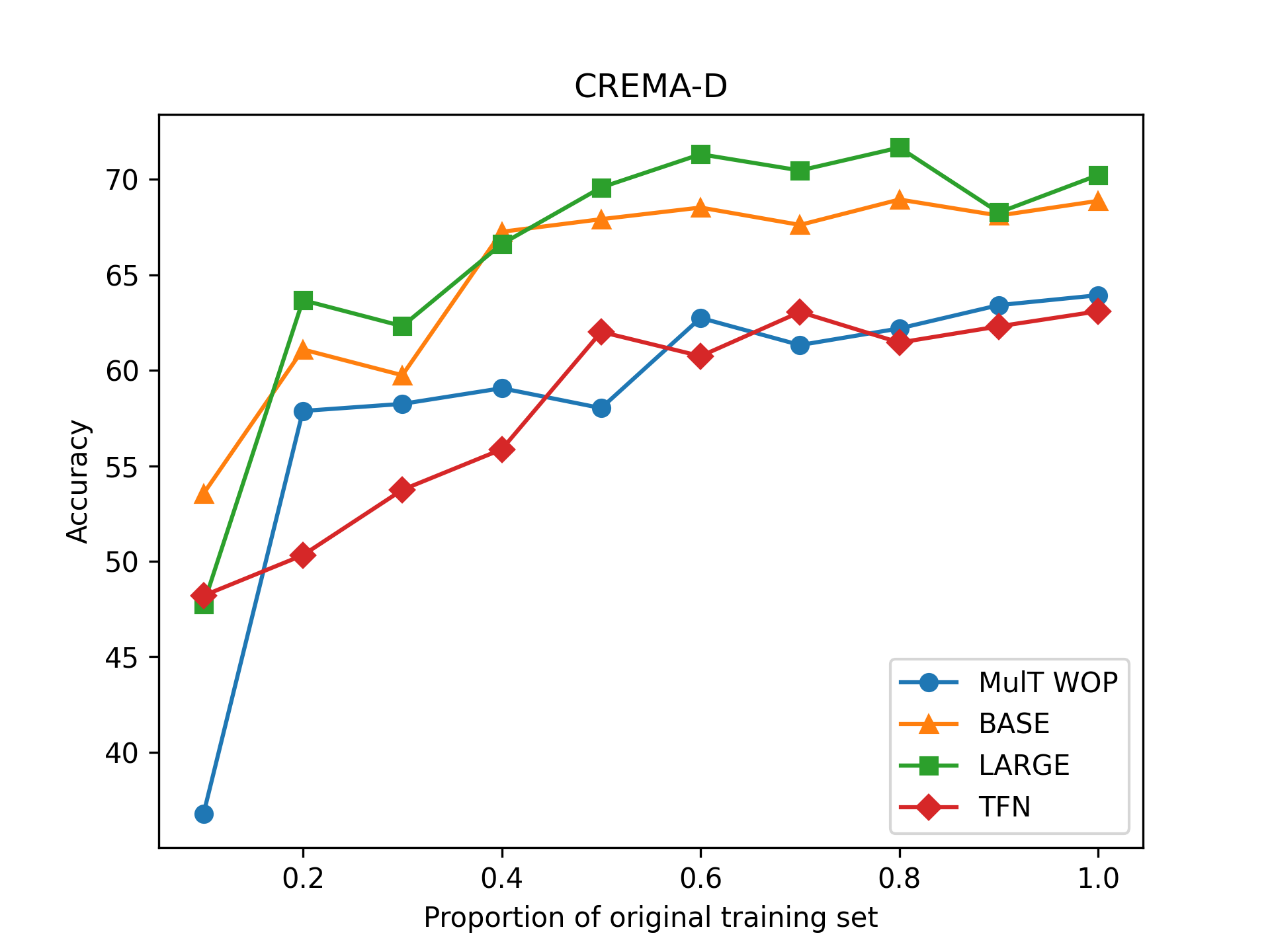}
\end{subfigure}
\vspace{-10pt}
     \begin{subfigure}[b]{.8\columnwidth}
  \includegraphics[width=\linewidth]{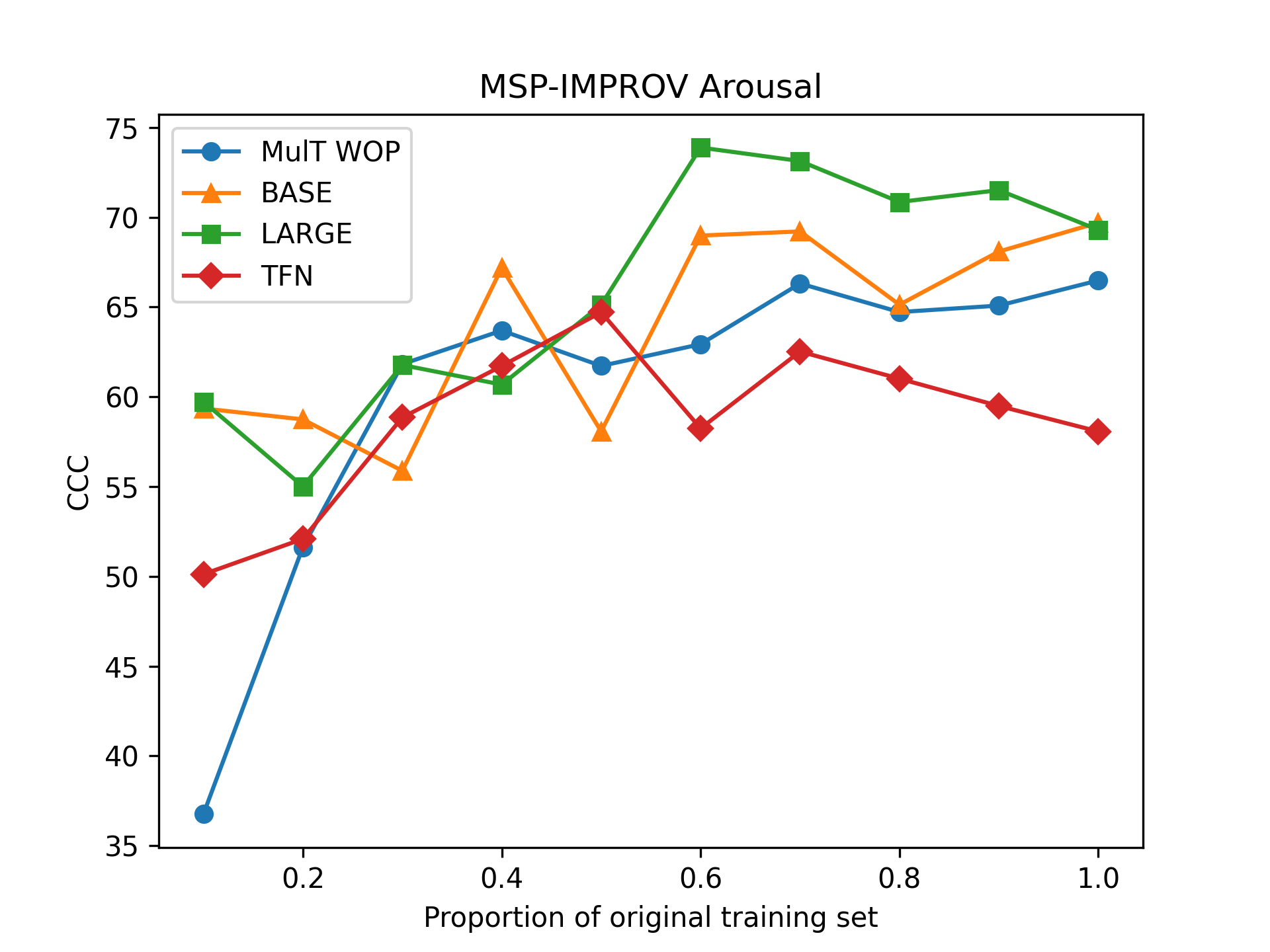}
\end{subfigure}
\vspace{-10pt}
     \begin{subfigure}[b]{.8\columnwidth}
  \includegraphics[width=\linewidth]{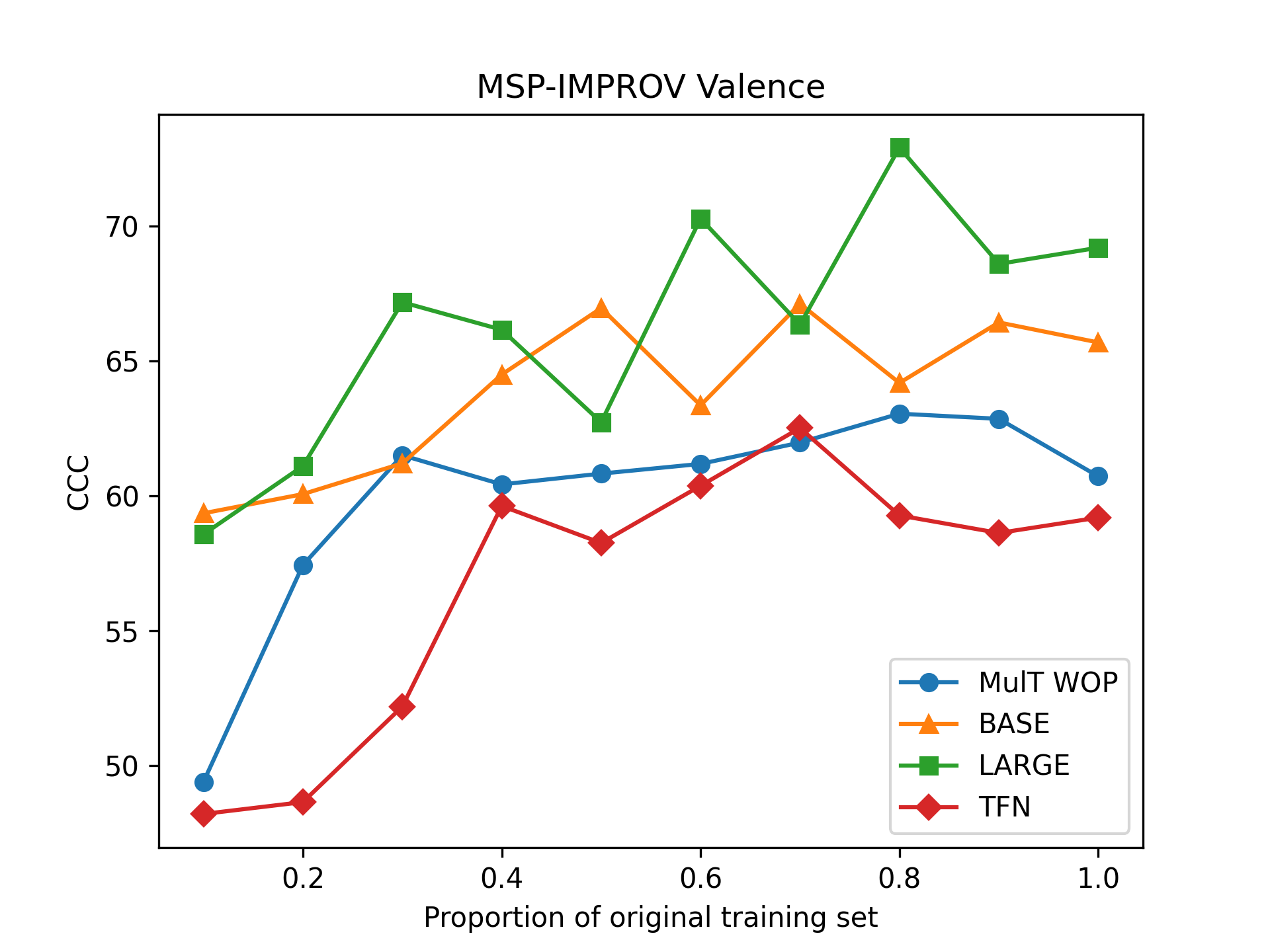}
\end{subfigure}
\caption{Performance of the models with restricted data.}
\label{fig:limited_data}
\vspace{-15pt}
\end{figure}

\subsection{Pretraining implementation details}
Following prior work on pretraining Transformers \cite{devlin2019bert, liu2020mockingjay}, we implement the pretraining task with two model settings: BASE and LARGE. For both configurations, we set the number of attention heads to $12$, the number of consecutive frames for masking to  $3$ ($\sim0.6$ sec) and the length of each processed sequence is $50$ ( $\sim10$ sec). 

The hidden sizes for each of the audio and visual modality are $288$ (BASE) and $576$ (LARGE). The sizes of the feed-forward layers in each cross-modal attention block are $1152$ (BASE) and $1536$ (LARGE). The sizes of the feed-forward layers in each Self-Attention Block are $2304$ (BASE) and $3072$ (LARGE). The BASE configuration has 6 $A\rightarrow V$ cross-modal attention blocks, 6 $V\rightarrow A$ cross-modal attention blocks and 6 self-attention blocks, which sums up to 38.3M parameters. The LARGE configuration has 8 $A\rightarrow V$ cross-modal attention blocks, 8 $V\rightarrow A$ cross-modal attention blocks and 8 self-attention blocks that totals 89.2M parameters. We train both models with the Adam optimizer \cite{kingma2014adam}. The learning rate is set to $5e^{-4}$, with a linear learning rate scheduler and a warmup portion of $0.1$. Both models are trained with a batch size of $64$ for $30$ epochs.

\subsection{Application on downstream task}

For fine-tuning, the last elements from the outputs of the pretrained MulT are passed to a residual block followed by a fully-connected layer to make final predictions. We compare the performance of the fine-tuned MulT to 4 baseline models: Early Fusion GRU (EF-GRU), Late Fusion GRU (LF-GRU), the Tensor Fusion Network (TFN) \cite{zadeh2017tensor} and the Multimodal Transformer without the pretrained weights initialization. It is important to note that TFN only processes static inputs, \textit{i.e.}, each modality of a sample is represented by a vector. However, we decide to include it as a baseline model because TRILL is originally developed to represent an audio as a whole with a vector \cite{shor2020towards}. Hence, for each video, we use the vector representation extracted from TRILL for the acoustic modality  along with the average of the 17 OpenFace AU intensities for the visual modality as inputs to TFN. 

To make the comparisons fair for EF-GRU and LF-GRU, we make them Bidirectional and control the hidden size as well as number of layers such that the number of parameters of these models are approximately the same with the BASE configuration of MulT. For MulT without pretrained weights, we perform experiments with both the BASE and LARGE configurations and report the better performing configuration based on the validation set. We train all of the models until early stopping occurs on the validation set. 

Table \ref{tab:comparisons} shows the performance of different models on the CREMA-D and MSP-IMPROV datasets. Since CREMA-D's classes are balanced, we use accuracy as our evaluation metric. Following \cite{chao2019enforcing, atmaja2020deep, parthasarathy2017jointly}, we report the Mean Absolute Error (MAE) and the Concordance Correlation Coefficients (CCC) to assess the quality of the regression models on the MSP-IMPROV dataset. 

The fine-tuned models outperform the baseline models by a considerable margin. For emotion recognition accuracy, we see a 5\% improvement for the BASE model and 7\% improvement for LARGE model in comparison with the baselines. On the MSP-IMPROV dataset, the fine-tuned models also shows improvements over the baselines on both Arousal and Valence regressions. Specifically, fine-tuning the BASE model achieves $3.2\%$ and $5.1\%$ gain in CCC for Arousal and Valence regression respectively. Although there might be discrepancies between train-validation-test set split, we find our best results (accuracy of 70.22\% on CREMA-D, CCC of 0.697 and 0.692 on MSP-IMPROV Arousal and Valence regression) competitive with existing benchmarks on CREMA-D \cite{ghaleb2019metric, birhala2020temporal, ghaleb2020multimodal} and MSP-IMPROV \cite{atmaja2020multitask, atmaja2020deep}.

\subsection{Limited resource setting}
Since the ultimate motivation of transfer learning is to reduce the requirements on labeled data, we are interested in exploring the capability of the pretrained MulT in a limited resource setting. Figure \ref{fig:limited_data} shows the performance of the models when only $N\%$ of the original training set are used for training. We can see that the performance drop curves for the pretrained models are less steep in comparison with training MulT from scratch and TFN. With only $10\%$ of the original training set (less than $500$ training samples on both datasets), finetuning the pretrained models outperforms training from scratch by at least $10\%$ for emotion recognition, and more than $20\%$ and $10\%$ CCC improvements for Arousal and Valence regression respectively. This further suggests the robustness of the pretrained weights in preventing overfitting with limited data. We can also note that TFN tends to perform better than training MulT from scratch with small training sets, which is expected because light models tend to be less susceptible to overfitting than complex ones with limited data.

\section{Conclusion}
In this study, we present the potential of pretraining the Multimodal Transformer architecture \cite{tsai2019multimodal} to model human communicative behaviors. We validate the usefulness of the pretrained model for the task of emotion recognition on two datasets, and demonstrate the robustness of the model in a low-resource setting. In the future, we will explore the performance of the model on other domains relating to communication such as mental health assessment. 

\section*{Acknowledgment}
Research was sponsored by the Army Research Office and was accomplished under Cooperative Agreement Number W911NF-20-2-0053. The views and conclusions contained in this document are those of the authors and should not be interpreted as representing the official policies, either expressed or implied, of the Army Research Office or the U.S. Government. The U.S. Government is authorized to reproduce and distribute reprints for Government purposes notwithstanding any copyright notation herein.

\bibliographystyle{IEEEbib}
\bibliography{refs}

\end{document}